# Development of Si based anodes for Li-ion batteries from a rational component design

Keke Chang[1], Yong Du[2]

[1]*Key Laboratory of Marine Materials and Related Technologies, Zhejiang Key Laboratory of Marine Materials and Protective Technologies, Ningbo Institute of Materials Technology and Engineering, Chinese Academy of Sciences, Ningbo, Zhejiang 315201, China*

[2]*State Key Laboratory of Powder Metallurgy, Central South University, 410083, Changsha, Hunan, China*

Email: changkeke@nimte.ac.cn

## Abstract

Inspired by the wisdom of metallurgists in designing new alloys, the Integrated Computational Materials Engineering (ICME) based design strategy is proposed for development of Si based anodes for Li-ion batteries (LIBs). The strategy starts with a rational component design of Si-$X$, where $X$ is the additive component(s) helping to overcome the problems of the pure Si anodes. An optimization of the composition, structure, property and performance of the Si-$X$ anode is followed to fulfill the requirements for its commercialization. In addition to the widely applied designing scheme for the nanostructured Si anodes, the presently proposed one from the ICME based rational component design is expected to accelerate the discovery of the promising Si based anodes for commercial LIBs.

## Keywords

A Li-ion battery (LIB), typically with an anode of graphite and a cathode of Lithium transition oxide, is widely used for portable electronic devices [1-5]. Because the demand for longer lasting rechargeable batteries with higher power and larger capacity in electric vehicles is ever increasing, significant scientific and engineering efforts have been made to develop novel anodes to replace graphite. Silicon (Si) is one of the most attractive LIB anode materials, owing a factor of ~10 larger capacities than graphite and superior safety compared to graphite [2,6]. Meanwhile, Si is abundant, cheap and environmentally friendly. However, three major challenges block the application and commercialization of Si as an LIB anode: large volume expansion upon Li incorporation (> 300 %), instability of the solid electrolyte interface (SEI) layer and its low electrical conductivity [2,6], as summarized in Figure 1.

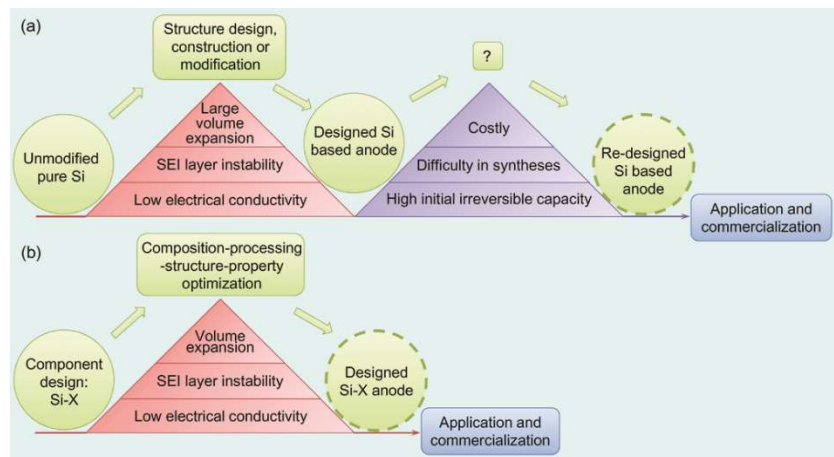

**Figure 1.** Schematic strategies of developing novel Si based anodes for commercial Li-ion batteries: (a) generally reported one following structure design, construction or modification, which was typically applied for nano-structure; and (b) the presently proposed one according to the Integrated Computational Materials Engineering (ICME) based rational component design.

Various approaches have been employed to develop novel Si based anodes (see Ref. [1,2,6-8] and references therein). One approach is classified as design of Si structures in various dimensions (D), such as 0D (nanoparticle), 1D (nanowire and nanotube) and 2D (film) [1,6,9,10]. Another approach is the hierarchical structure construction including core-shell, yolk-shell, embedding and so on. Moreover, modification of Si structures with porous, sandwiched or multi-layered features are effective ways to improve their performance. These approaches focus on tuning the Si structure, typically the nanostructure, and can be summarized as a strategy shown in Figure 1a. Consequently, significant progress has been achieved in the properties and performance of Si anodes. However, drawbacks emerge as new barriers for commercialization of these anodes [8,9], as illustrated in Figure 1a. Firstly, the nano-techniques may lead to the cost several times increased. Secondly, the sophisticated synthesis methods make mass production difficult. Thirdly, high surface area of the nanostructures results in high initial irreversible capacities. Accordingly, the Si based anodes require re-design prior to their practical application and commercialization for LIBs.

Alternatively, another strategy is proposed, namely the Integrated Computational Materials Engineering (ICME) based rational component design. Materials research using the trial-and-error method is extremely effort demanding and time consuming [10-12]. ICME, involving quantitative high-throughput and/or multi-scale computation supported by key experiments, has been raised as the potential solution to design future anodes through pre-selection and optimization of candidate materials at an unprecedented rate with a relatively low cost [13-15]. *Ab initio* (or first-principles) method is widely applied to calculate electronic structures of materials at the atomic scale [16-18], which can help to understand magnetic, thermodynamic, elastic and many other properties of Si based anodes. The physical movements of Li in LIBs can be studied through molecular dynamics (MD) simulation [6,19]. The CALPHAD (CALculation of PHAse Diagrams) approach is used to obtain phase diagrams by assessing Gibbs energy functions of phases [20,21]. The phase-field modeling (PFM) approach is utilized to model microstructure evolution and the finite element method (FEM) is popular in property and performance predictions for materials at a macro-scale, acting as a guidance for processing and manufacturing [19]. Through the integrated methods (*ab initio*, CALPHAD, MD, FPM, FEM, etc.), ICME would provide a significant insight into the quantitative composition-structure-property relationships in the complex Si based systems from the quantum- to macro-scale and thereby accelerate the design of new anodes.

A first major step following ICME to develop novel Si based LIB anodes is the rational component design of Si-*X*, as shown in Figure 2b. *X* is the additive component(s) to form Si-*X* solid solutions or compounds, which would help to overcome the three problems of the pure Si anodes. For instance, Al, C or Cu addition enhances the electrical conductivity, while transition metals can buffer the volume expansion of Si. The feasibility of the strategy is foreseen from the reported Si-*X* as promising anodes, just to mention a few, Si-C, Si-C-O, metallic Si-*M* (*M* = Al, Mg, Ca, Fe, Ni, Cu), Si-Al-*M* (*M* = V, Cr, Mn, Fe, Co, Ni), Si-Ni-Ti (see Ref. [2,6,9,22] and references therein). All these anodes were developed with support of *ab initio*, MD or CALPHAD results. Particularly, thermodynamic calculations and phase diagrams were regarded as the basis for metallic element selection [23]. Note that *X* addition should affect other battery performances (such as battery capacity, power density, safety and so on) not too much, since the commercialization of the Si-*X* anode is a compromise of its overall properties.

As the next step, the composition, processing, structure, property and performance of the Si-*X* anode would be optimized to fulfill the requirements for its commercialization. The basic consideration is that the behavior of an alloy is generally determined by its structure at different scales, while its structure is governed by its composition and the processing. Compared to the synthesis methods in the first strategy for nanostructures (Figure 1a), the economical and widely-applied metallurgy, melt-spinning and thin film techniques [9,19] are usually performed in the second strategy. Hence, owing to the refined process for mass production, the designed Si-*X* anodes will be immune to the three issues emerging from the nanostructure based design following the first strategy. Note that the strategy is an alternative but not a short cut, the appearing "one-step" design in Figure 1b also requires a systematic and long-term study of the Si-*X* system for a commercialized LIB anode.

Interestingly, the presently proposed strategy was also inspired by the wisdom of metallurgists in designing new alloys (e.g., development from pure Fe to steel [26-28], from pure Ti to Ti alloys [29-31] and from pure Al to Al alloys [32,33]), which has been thriving for over a century. Both fundamental and applied researches are needed to realize the application of the strategy [1,34,35]. Simultaneously, apart from the above-mentioned approaches, high-throughput experimentation and data-mining techniques will help to obtain LIBs' information about battery capacity, Li ion diffusion, charge transfer, structure evolution, and so on. ICME is being proven to be reliable for novel materials development. In addition to the widely applied designing scheme for the nanostructured Si anodes, the presently proposed one from the ICME based rational component design is expected to accelerate the discovery of the promising Si based anodes for commercial LIBs.

**Funding**

This work was supported by the National Natural Science Foundation of China (51701232) and CAS Pioneer Hundred Talents Program.